\documentstyle[prl,aps,twocolumn,draft,float]{revtex}

\input epsf.sty

\begin{document}

\draft

\wideabs{

\title{Metamagnetism and critical fluctuations in high quality single
crystals of the bilayer ruthenate Sr$_3$Ru$_2$O$_7$}

\author{R. S.  Perry, L. M.  Galvin, S. A.  Grigera, L.  Capogna, 
A. J. Schofield, A. P.  Mackenzie}

\address{School of Physics and Astronomy, University of Birmingham, Edgbaston,
Birmingham B15 2TT, United Kingdom.}

\author{M. Chiao and S. R.  Julian}

\address{Cavendish Laboratory, University of
Cambridge, Madingley Road, Cambridge CB3 OHE, United Kingdom.}

\author{S. Ikeda$^{1,2}$, S. Nakatsuji$^2$ and Y. Maeno$^{2,3}$}

\address{$^1$Electrotechnical Laboratory, Tsukuba, Ibaraki 305--8568,
Japan.\\ $^2$Department of Physics, Kyoto University, Kyoto
606--8502, Japan.\\ $^3$CREST, Japan Science and Technology
Corporation, Kawaguchi, Saitama 332--0012, Japan.}

\author{C. Pfleiderer}

\address{University of Karlsruhe, Institute of Physics, Kaiserstr.
12, D-76128 Karlsruhe, Germany.}

\maketitle

\begin{abstract}
We report the results of low temperature transport, specific heat and
magnetisation measurements on high quality single crystals of the
bilayer perovskite ${\rm Sr_3Ru_2O_7}$, which is a close relative of
the unconventional superconductor ${\rm Sr_2RuO_4}$.  Metamagnetism is
observed, and transport and thermodynamic evidence for associated
critical fluctuations is presented.  These relatively unusual
fluctuations might be pictured as variations in the Fermi surface
topography itself.  No equivalent behaviour has been observed in the
metallic state of ${\rm Sr_2RuO_4}$.

\end{abstract}


}

Research over the past decade has shown the potential of perovskite
ruthenate metals to play a pivotal role in our understanding of the
behaviour of strongly correlated electrons.  The position of the Fermi
level in bands resulting from the hybridisation of oxygen 2p and
ruthenium 4d levels leads to ground state behaviour covering a wider
range than that seen in almost any other transition metal oxide
series.  Pseudocubic SrRuO$_3$ is a rare example of an itinerant
ferromagnet based on 4d electrons, and has a good lattice match to the
cuprates \cite{1,2}.  ${\rm Sr_2RuO_4}$ has the layered perovskite
structure with a single RuO$_2$ plane per formula unit.  It is
strongly two-dimensional, and shows a Pauli-like paramagnetic
susceptibility \cite{3}.  It is best known for its unconventional
superconductivity \cite{3}, which is thought to involve spin triplet
pairing \cite{4}.  Structural distortions in Sr-based ruthenates are
either small or absent, but substituting Ca for Sr introduces larger
rotations of the Ru--O octahedra, causing bandwidth narrowing and
changes to the crystal field splitting.  Thus, although Ca and Sr are
both divalent cations, the properties of the Ca-based materials are
markedly different.  CaRuO$_3$ is a paramagnetic metal with a large
mass enhancement \cite{5}, while Ca$_2$RuO$_4$ is an antiferromagnetic
insulator \cite{6}.  This diversity shows that the ruthenates are
characterised by a series of competing, nearly degenerate
instabilities, giving a clear motivation for the careful investigation
of all the compounds in the series.  An even more important feature of
the ruthenates is that, in contrast to 3d oxides such as most
manganites and many cuprates, no explicit chemical doping is required
to produce metallic conduction.  This gives a unique opportunity to
probe a wide range of correlated electron physics in the low disorder
limit, leading to considerable advances in understanding. The
superconductivity of ${\rm Sr_2RuO_4}$, for example, is strongly
disorder-dependent \cite{7}, and further examples of unconventional
superconductivity may be expected in other ruthenates if they can be
grown with mean free paths as long as those of ${\rm Sr_2RuO_4}$.  Of
particular interest is the subject of this study, ${\rm Sr_3Ru_2O_7}$,
which has a Ru--O bilayer per formula unit, and hence an effective
dimensionality which is intermediate between those of ${\rm
Sr_2RuO_4}$ and SrRuO$_3$.

The synthesis of ${\rm Sr_3Ru_2O_7}$ in polycrystalline form has been
reported by several groups over the past three decades \cite{8,9,10},
but investigations of its electrical and magnetic properties were not
carried out until the past few years.  There has been some variation
in the reports and interpretation of its ground state.  Cava and
co-workers showed that the magnetic susceptibility ($\chi$) of powders
obeyed a Curie-Weiss law with a maximum in the susceptibility at
approximately 20K and a negative Weiss temperature, which they
interpreted in a local moment picture as antiferromagnetism \cite{11}.
Early reports on single crystals grown from a SrCl$_2$ flux in Pt
crucibles, however, gave evidence for weak itinerant ferromagnetism
\cite{12}.  Elastic neutron scattering measurements on polycrystalline
materials showed no sign of either kind of long-range order \cite{13}.

The lack of long-range order was confirmed by Ikeda and co-workers
\cite{14}, who succeeded in growing much purer crystals (residual
in-plane resistivity, $\rho_{\rm res}$, of 3--4 $\mu\Omega$cm) in an
image furnace.  The magnetic susceptibility of these crystals
reproduces the basic features of the paramagnetic susceptibility
reported in ref. 11, with a pronounced maximum at approximately 16K.
Although the resistivity has an anisotropy ratio of several hundred,
the susceptibility is much more isotropic, with an anisotropy at 16K
of only $20\%$.  At lower temperatures, below 5K, it becomes
completely isotropic and temperature independent.  The low temperature
value of $\chi = 1.5 \times 10^{-2}$emu/mol--Ru is over an order of
magnitude larger than that of ${\rm Sr_2RuO_4}$, while values between
1.5 and 2.5 times larger have been reported for the electronic
specific heat per Ru mole \cite{14,15}. Combined with the observation
of a $T^2$ dependence of the resistivity ($\rho$) in the same
temperature range, the data strongly suggest that ${\rm Sr_3Ru_2O_7}$
enters a Fermi liquid metallic state at low temperatures.  Published
electronic band structure calculations predict a Fermi surface which
is very similar to that of ${\rm Sr_2RuO_4}$ \cite{16}, giving no
expectation of substantial differences between the properties of the
two materials.

Here, we report the results of low temperature transport, magnetic and
thermodynamic measurements on the crystals studied in ref. 14 and some
even cleaner ones ($\rho_{{\rm res}} = 2~ \mu \Omega {\rm cm}$).  We
conclude that the low temperature properties of ${\rm Sr_3Ru_2O_7}$
are strongly influenced by critical fluctuations associated with
metamagnetism, something that has never been observed in ${\rm
Sr_2RuO_4}$.
\begin{figure}
\centerline{\epsfxsize=8.5 cm \epsfbox{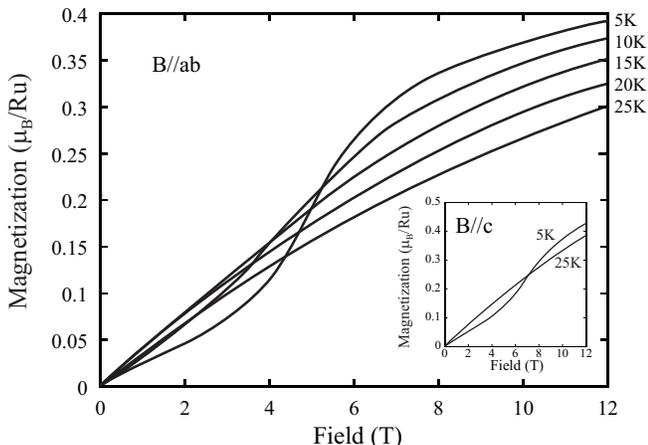}}
\caption[fig1]{The magnetisation of single crystal ${\rm Sr_3Ru_2O_7}$
for magnetic fields applied in the $ab$ plane.  The high field data
drop monotonically, with the highest value seen at 5K.  Metamagnetism
is clearly seen in the 5K and 10K data, at a field of approximately
5.5 tesla.  No anisotropy is observed for different field directions
within the ab plane (data not shown).  Inset: Data at 5K and 25K with
the magnetic field applied along the $c$ axis.  For this orientation
the metamagnetic field is approximately 7.7 tesla (see Fig. 2).}
\label{fig 1}
\end{figure}

Several growth runs of single crystals of ${\rm Sr_3Ru_2O_7}$ were
performed crucible-free in an image furnace in Kyoto. The resistivity
of over fifty pieces taken from three growth rods was studied down to
4K using standard low frequency ac methods in a small continuous flow
cryostat.  Residual resistivities ranging between $2~ \mu \Omega {\rm
cm} $ and $15~ \mu \Omega {\rm cm} $ were observed, with most samples
from the latter batches toward the high purity end of the range.  Six
of these were used for further study.  Magnetisation measurements were
performed down to 5K in a commercial vibrating sample magnetometer.
Magneto-transport was studied at ambient pressure in both $^4$He
systems and a dilution refrigerator in Birmingham, and in a $^4$He
system in Karlsruhe.  Specific heat measurements were performed in
Kyoto using commercial magneto-thermal apparatus.

Our magnetisation results for ${\rm Sr_3Ru_2O_7}$ are summarised in
Fig. 1.  For magnetic fields applied in the $ab$ plane, a rapid
super-linear rise in the magnetization is seen with a characteristic
field of approximately 5.5 tesla at 5K.  Such behaviour can be
described as metamagnetism (for further discussion see below).  As the
temperature is raised, the metamagnetism disappears.  The data give no
evidence for it moving to higher field; the dominant effect is a
broadening which makes it impossible to define.  We have checked
extensively for anisotropy with respect to the direction of the
in-plane field, but none is observed within our resolution.  There is
some anisotropy if the field is applied along $c$, as shown in the
inset.  Although data were also taken at 10, 15 and 20K, they lie
sufficiently close to confuse the plot, so have been omitted for
clarity.

In order to obtain information on the metamagnetism to lower
temperatures, we have studied its effects on the magnetoresistance
(MR).  Field sweeps to 14 tesla were performed between 50 mK and 20K
for three standard configurations of the in-plane MR $\rho : B
\parallel c,~ I \parallel ab;\ B \parallel I \parallel ab$ and $(B
\perp I) \parallel ab$.  The results are summarised in Fig.  2, in
which a small subset of representative data is presented.  For $B
\parallel c, I \parallel ab$, the weak-field MR is quadratic in $B$.
At higher fields, the metamagnetism is clearly seen in the MR.  For $T
\geq$ 5K, the width of the feature is similar to that seen in the
magnetisation, but at low temperatures it sharpens considerably,
although there is evidence for extra structure at 11 tesla.  For $B
\parallel ab$ there is clear evidence of a split transition for both
$B \parallel I$ and $B \perp I$.  The transport measurements show that
the metamagnetic transition field is essentially temperature
independent for any direction of the applied field.

The term metamagnetism can be applied to qualitatively different
physical phenomena.  In insulators, it describes changes from ordered
antiferromagnetic states at low field to ferromagnetically polarized
states at high field via `spin-flip' or `spin-flop' processes
\cite{17}.  In metallic systems such as ${\rm Sr_3Ru_2O_7}$, the
change in magnetization is due to a rapid change from a paramagnetic
state at low fields to a more highly polarized state at high fields,
via either a crossover or a phase transition.  Although the
distinction between a crossover and a phase transition can be an
important one, in this case it is less relevant, because some of the
basic physics is common to both.  Firstly, the observation of
itinerant metamagnetism demonstrates the existence of strong
ferromagnetic coupling in the system.  Secondly, low temperature
critical points are likely in either scenario.  A crossover might be
linked to close proximity in phase space to a quantum critical point
that can be reached only by the application of some other control
parameter.  If the metamagnetism is due to a $T=0$ phase transition
along the magnetic field axis, the transition would be expected to be
first order (since it is like a density transition of spins in the
presence of a symmetry-breaking field).  However, the first order
transition line might terminate in a critical point at very low
temperatures.  A natural question, then, is whether there is evidence
for critical fluctuations associated with the metamagnetism in ${\rm
Sr_3Ru_2O_7}$.  To address this issue, we have performed measurements
of the temperature-dependent resistivity and specific heat.
\begin{figure} 
\centerline{\epsfxsize=7cm \epsfbox{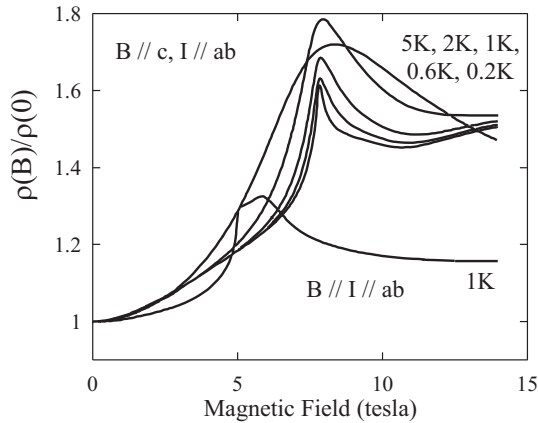}}
\vspace{1em}
\caption[fig2]{The magnetoresistance (MR) of single crystal
Sr$_3$Ru$_2$O$_7$ at a series of temperatures below 1K.  For $B
\parallel c$, the very broad peak at 5K (comparable to the total width
of the feature seen in M(B)) sharpens considerably at low
temperatures.  For in-plane fields, the low temperature MR gives
evidence for some peak splitting, as shown by data at 1K for this
orientation.}
\label{fig 2}
\end{figure}

In Fig. 3 we show a colour plot of the power-law behaviour of the
resistivity of ${\rm Sr_3Ru_2O_7}$ as a function of temperature and
field, for $I \parallel ab \parallel B$.  At low and high field, the
quadratic temperature dependence expected of a Fermi liquid is seen at
sufficiently low temperature, but near the metamagnetic field, the
Fermi liquid region is suppressed to below our lowest temperature of
measurement (2.5K for these precise temperature sweeps).  These data
are strongly suggestive of the existence of a critical point at a
temperature low on the scale of the measurement temperature, at a
field close to the metamagnetic field.  They do not constrain it to
lie exactly on the field axis, but the applied field is clearly
pushing the system very close to it, and its associated critical
fluctuations are playing a central role in determining the properties
of the metallic state.
\begin{figure} 
\centerline{\epsfxsize=7cm \epsfbox{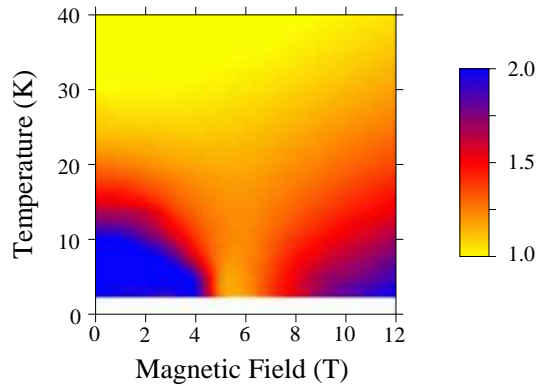}}
\vspace{1em}
\caption[fig3]{The evolution of the temperature dependence of the
resistivity of ${\rm Sr_3Ru_2O_7}$ as a function of temperature with
$B \parallel I \parallel ab$.  The plot was constructed by
interpolating the results of temperature sweeps at fourteen values of
applied field (0 to 12 tesla in 1 tesla steps, with an additional
sweep at 5.5 tesla). The colours show the change in the parameter
$\alpha$ for a temperature-dependent resistivity of the form
$T^\alpha$ .  The $T^2$ behaviour expected of a Fermi liquid is
observed at low temperatures only for fields well away from the
metamagnetic field of approximately 5.5 tesla identified from Fig. 2.}
\label{fig 3}
\end{figure}

Measurement of the specific heat gives further experimental support
for the existence of critical fluctuations.  Data for the electronic
specific heat ($C_{\rm el}$) divided by temperature are shown in
Fig. 4.  In zero field, $C_{\rm el} / T$ (a measure of the
quasiparticle mass) rises as the temperature falls below 20K, but then
crosses over at lower temperatures on a scale similar (but not
identical) to the crossover to the quadratic dependence of the
resistivity seen in Fig. 3.  The application of fields parallel to the
$c$ axis suppresses this crossover, so that $C_{\rm el} / T$ rises
steeply at low temperatures.  At 9 tesla there is evidence for the
divergence being cut off at low temperatures, but at 7.7 tesla, no
cut-off is seen.  As well as giving thermodynamic support for the
critical fluctuations suggested by the transport measurements, the
specific heat data emphasise the closeness of the link between the
metamagnetism and these fluctuations.  If the critical point
dominating the fluctuations seen in Fig. 3 were primarily controlled
by some parameter other than the field, then changing the field
orientation would be expected to make very little difference.  Here,
however, we have tuned into the divergence of $C_{\rm el} / T$ by
cooling at the metamagnetic field for $B \parallel c $ (identified as
7.7 tesla by the peak in Fig. 2) rather than the value of
approximately 5.5 tesla that would be deduced from Figs. 2 and 3 for
$B \parallel ab$.

We believe that the above data give good evidence for a close
association between metamagnetism and critical fluctuations in ${\rm
Sr_3Ru_2O_7}$.  The best way to view itinerant metamagnetism is that
the extra moment is generated by a polarization of the up and down
spin Fermi surfaces, so our results suggest that ${\rm Sr_3Ru_2O_7}$
is an excellent material in which to study a very interesting form of
low temperature fluctuation, that of the Fermi surface itself.
Although this is a ${\bf q}=0$ picture, the polarized and unpolarized
Fermi surfaces are likely to have different nesting properties in a
nearly two-dimensional material, so a coupling with fluctuations at
higher ${\bf q}$ is to be expected.
\begin{figure} 
\centerline{\epsfxsize=8cm \epsfbox{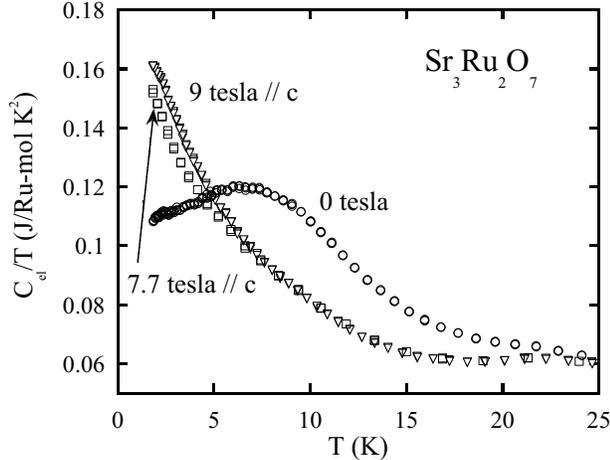}}
\caption[fig4]{The electronic specific heat divided by temperature for
${\rm Sr_3Ru_2O_7}$ at low temperatures, for magnetic fields applied
parallel to $c$. Subtracting the $T^3$ lattice contribution gave a
Debye temperature of 360K, in good agreement with previous work on
polycrystalline samples \cite{15}. Apparently divergent behaviour is
seen when the field is tuned to the metamagnetic peak for this field
orientation, identified from Fig. 2 to be 7.7 tesla.  If the field is
increased to 9 tesla, the divergence is cut off.}
\label{fig 4}
\end{figure}

Itinerant electron metamagnetism has previously been observed in
several systems (notable recent examples are MnSi \cite{18} and
CeRu$_2$Si$_2$ \cite{20}), but this is, to our knowledge, the first
time that such clear evidence has been seen for a magnetic field
driving a system closer to criticality.  The fact that it has been
seen in a close structural relative of ${\rm Sr_2RuO_4}$ gives the
result further significance, and provides a second motivation for more
detailed investigation of the spin fluctuations in ${\rm
Sr_3Ru_2O_7}$.  There is good and growing evidence for triplet
superconductivity in ${\rm Sr_2RuO_4}$, but, so far, little
experimental evidence for the strong low--{\bf q} fluctuations that
might na\"{\i}vely be expected to give the underlying binding
mechanism.  The only structure that has been observed is a nesting
feature at (0.6 $\pi/a$, 0.6 $\pi/a$) \cite{21}, and no metamagnetism
has been observed in fields of up to 33 tesla \cite{22}.  In ${\rm
Sr_3Ru_2O_7}$, our results in combination with those of ref. 14
clearly demonstrate the existence of a low ${\bf q}$ enhancement, but
no superconductivity has been observed down to our best residual
resistivity of 2 $\mu \Omega {\rm cm}$.  It will be very interesting
to see if there is a significant enhancement at high ${\bf q}$ as
well, and to investigate the underlying reasons for the pronounced
differences in the magnetic properties between the two materials.  In
this respect, the orthorhombicity recently reported in ${\rm
Sr_3Ru_2O_7}$ \cite{23} may prove to be significant.  More generally,
continued work on ${\rm Sr_3Ru_2O_7}$ is likely to play an important
role in efforts to understand unconventional superconductivity in the
ruthenates.  A search for quantum oscillations, an investigation of
the effect of pressure on the metamagnetism and inelastic neutron
scattering measurements are all desirable, in combination with further
improvements in sample purity.

In conclusion, we have presented evidence that the low temperature
properties of ${\rm Sr_3Ru_2O_7}$ are strongly influenced by critical
fluctuations associated with the presence of metamagnetism.  Our
observations highlight important and somewhat unexpected differences
with the metallic state that exists in its well-studied structural
relative, the unconventional superconductor ${\rm Sr_2RuO_4}$.

We are grateful to C. Bergemann, N. R. Cooper, V. J. Emery,
E. M. Forgan, H. von Loehneysen, G. G. Lonzarich, A. J. Millis,
A. Rosch, and D. J. Singh for fruitful discussions.  This work was
supported by the EPSRC (UK), the Royal Society, the Leverhulme Trust,
NSERC (Canada), CREST-NST (Japan) and ESF (FERLIN Collaboration).

\end{document}